\documentclass[showpacs,preprintnumbers,
amsmath,amssymb,aps,prd,nofootinbib,superscriptaddress,citeeqsecnum,notitlepage,11pt]{revtex4-1}
\usepackage[T1]{fontenc}
\usepackage{amsmath,xcolor}
\usepackage{amsmath, amssymb,ascmac,fancybox,mathrsfs}
\usepackage{color,booktabs,fancyhdr,bm,comment,cancel}
\usepackage{hyperref}
\usepackage{graphicx,graphics}

\makeatletter
\renewcommand{\p@subsection}{}
\makeatother

\makeatletter
\newcommand{\xequal}[2][]{\ext@arrow 0055{\equalfill@}{#1}{#2}}
\def\equalfill@{\arrowfill@\Relbar\Relbar\Relbar}
\makeatother

\renewcommand{\thesection}{\arabic{section}}
\renewcommand{\thesubsection}{\arabic{section}.\arabic{subsection}}

\makeatletter
\renewcommand{\theequation}{\arabic{section}.\arabic{equation}}
\@addtoreset{equation}{section} 
\makeatother

\newcommand{\chushi}[1]{ }

\newcommand{\angleN}[1]{ \langle #1 \rangle }

\newcommand{\roundLR}[1]{ \left( #1 \right) }
\newcommand{\roundN}[1]{ ( #1 ) }

\newcommand{\roundBB}[1]{ \Biggl( #1 \Biggr) }

\newcommand{\curlyN}[1]{ \{ #1 \} }

\newcommand{\squareLR}[1]{ \left[ #1 \right] }
\newcommand{\squareN}[1]{ [ #1 ] }

\newcommand{\absLR}[1]{ \left| #1 \right| }

\newcommand{\im}{\text{Im}}

\newcommand{\diag}{\text{diag}}

\let\calccommentout\iffalse 
\let\calcshow\iftrue

\newcommand{\eq}[1]{\begin{equation}\begin{split} #1 \end{split}\end{equation}}

\newcommand {\mathsym}[1]{{}}
\newcommand {\unicode}[1]{{}}

\begin{document}

\title{Proper effective temperature of nonequilibrium steady state}

\author{Hironori Hoshino}
\affiliation{Department of Physics, Indian Institute of Technology Ropar, Rupnagar, Punjab 140 001, India}

\author{Shin Nakamura}
\affiliation{Department of Physics, Chuo University, Tokyo 112-8551, Japan}

\begin{abstract}
We define a proper effective temperature for relativistic nonequilibrium steady states (NESSs).
A conventional effective temperature of NESSs is defined from the ratio of the fluctuation to the dissipation.
However, NESSs have relative velocities to the heat bath in general, and hence the conventional effective temperature can be frame dependent in relativistic systems.
The proper effective temperature is introduced as a frame-independent (Lorentz invariant) quantity that characterizes NESSs.
We find that the proper effective temperature of NESSs is higher than the proper temperature of the heat bath in a wide range of holographic models even when the conventional effective temperature is lower than the temperature of the heat bath.
\end{abstract}

\maketitle

\section{Introduction}
Nonequilibrium steady state (NESS) is important in the study of nonequilibrium physics.
To realize a NESS, we need three ingredients: a heat bath, a system we study, and an external force.
The system is in contact with the heat bath and is driven to nonequilibrium by an external force.
When the external force is constant, we obtain a NESS.
In NESSs, the energy given by the external force is constantly dissipated into the heat bath, and the steady state is maintained.
In general, a NESS has a relative velocity to the heat bath.

One physical parameter that characterizes the NESSs is the effective temperature \cite{Cugliandolo:2011,Berthier:2002,Hayashi:2004}.
The effective temperature is defined by the relationship between the fluctuations and the dissipation, hence it has information about the nonequilibrium dynamics.
In relativistic systems such as the quark-gluon plasma or the astrophysical jets, the effective temperature may be affected by an effect of Lorentz transformation.
In this work, we consider the Lorentz transformation of effective temperature.
We define the ``proper effective temperature" as a Lorentz invariant quantity, which characterizes NESSs in a frame-independent way.

There is an example where the relativistic effect is not negligible in the effective temperature.
We consider a conformally invariant system, studied in \cite{Gubser:2006nz,Nakamura:2013yqa,Hoshino:2014nfa}, where a heavy object is moving in a heat bath.
Let the object be dragged at constant velocity $\vec v$ by a constant external force.
This is an example of NESS.
The effective temperature of this NESS, $T_*$, is given by $T_*=(1- v^2/c^2)^{1/4}T_0$ \cite{Gubser:2006nz,Nakamura:2013yqa,Hoshino:2014nfa}, where $T_0$ denotes the proper temperature of the heat bath.
When the velocity is small enough, the difference between $T_*$ and $T_0$ is approximately given by $T_* - T_0 \sim - v^2/(4c^2) \times T_0$.
This difference should include effects of interaction between the test object and the heat bath. 
If the difference also includes the relativistic effect, it should be of the same order of the relativistic correction $O(v^2/c^2)$.
We should examine the amount of the relativistic correction. 

Let us briefly review the study of ``Lorentz transformation" of temperature in equilibrium system, which has a long history, e.g.  \cite{Balescu:1968,Dunkel:2009tla,Farias:2017,Einstein:1907,Planck:1908,Landsberg:1966,Kampen:1968,Ott:1963,Landsberg:1980,Kaniadakis:2002zz,Kaniadakis:2005zk,Israel:1976tn} and references therein.
For finite volume systems at equilibrium, three different definitions of temperature
were proposed as follows.
Suppose that two observers observe the temperature;
the observer $O_0$ in the rest frame of the equilibrium system, and the observer $O$ moving at constant velocity $-\vec{v}$ relative to $O_0$.
In general, the observers $O_0$ and $O$ may measure different values $T_0$ and $T$ as the temperature, respectively, and these should be related by the Lorentz transformation. 
Historically, three different definitions $T_\ell=\gamma^\ell T_0 ~(\ell=\pm 1,0)$ were proposed, where $\gamma=1/\sqrt{1-v^{2}}$ is the Lorentz factor in the unit of $ c = 1$.
The transformation with $\ell=-1$ was proposed by Einstein \cite{Einstein:1907} and Planck \cite{Planck:1908}; that with $\ell=0$ was given by Landsberg \cite{Landsberg:1966} and van~Kampen \cite{Kampen:1968}; that with $\ell=1$ was proposed by Ott \cite{Ott:1963}\footnote{Possible ideas for experimental observation of the Lorentz transformation of temperature are also discussed in \cite{Landsberg:1980,Kaniadakis:2002zz,Kaniadakis:2005zk}.}.

All these descriptions are based on relativistic versions of the first law of thermodynamics.
The observer $O_0$ manifestly can employ the ordinary thermodynamics, while $O$ has several choices for definition of the relativistic thermodynamics.
The difference among the above temperatures originates from merely different choices of the relativistic definition of {\em heat} and {\em temperature} itself as follows.
\begin{itemize}
\item
For $\ell = -1$, the heat $\delta Q$ and $\delta Q_0$ for $O$ and $O_0$, respectively, are defined as they satisfy $\delta Q = \gamma^{-1} \delta Q_0$, and the temperature is defined by $T_{-1} dS \equiv	\delta Q$, where $S$ is the entropy of the system which is Lorentz invariant.
This formalism consequently yields $T_{-1}=T_0/\gamma$.
\vspace{1ex}
\item 
For $\ell = 0$, the thermal energy-momentum is defined as a four-vector version of the heat, $\delta Q^\mu$,
and the temperature is defined through $ T_0dS \equiv - u_\mu \delta Q^\mu$, where $u^\mu$ is the four-velocity of the system.
This formalism holds the temperature Lorentz invariant.
\end{itemize}
Note that the above formalisms are translatable into each other, and we can derive them from a conventional theory of local equilibrium (see also Appendix \ref{sec:local} for details).
Nevertheless, the formalism for $\ell=0$ and $-1$ help us to understand the physics in NESSs\footnote{For our purpose, we do not need the case of $\ell=+1$ (see Appendix \ref{sec:first_law} for details).}.

The characteristics of the two temperatures are as follows.
First, the Einstein-Planck's temperature ($\ell=-1$) is directly read from a distribution function, and hence this may be useful in experiments.
However, the ``energy" and the ``momentum" that appear in the Einstein-Planck's first law of thermodynamics do not form a four-vector (see Appendix \ref{sec:first_law} and \ref{sec:local}).
Secondly, the van Kampen's temperature ($\ell=0$) is Lorentz invariant, which is useful to describe phenomena without any dependence on observers.
Hence it is important in relativistic theory such as relativistic fluid dynamics.
This Lorentz invariant temperature is referred to as the proper temperature of the system.
In van Kampen's first law of thermodynamics, the energy and the momentum of a system are defined as a four-vector.

In NESSs, thermodynamics is not established. Hence we need another method to derive the Lorentz transformation of effective temperature.
In this work, we employ a conjecture called holography, which states the equivalence between a strongly-interacting quantum gauge theory and a classical gravity theory \cite{Maldacena:1997re,Gubser:1998bc,Witten:1998qj}.
A natural definition of temperature arises in the gravity dual without relying on thermodynamics: the Hawking temperature.
We find that the Hawking temperature corresponds to the Einstein-Planck's temperature.
Its Lorentz transformation is obtained from that of the dual geometry.

An advantage of holography is its applicability to nonequilibrium physics \cite{Hubeny:2010ry}.
Using holography, we can read the effective temperature of NESS as the Hawking temperature of an effective geometry \cite{Gursoy:2010aa, Sonner:2012if, Nakamura:2013yqa, Hoshino:2014nfa}. 
Therefore, the Lorentz transformation of the effective temperature is straightforwardly given by that of the effective geometry of the gravity dual.

In this work, the mean velocity of NESS that is phenomenologically defined in  \cite{Hoshino:2017air} is rediscovered as the velocity of the NESS relative to the heat bath.
We also define the ``proper effective temperature" and the ``rest frame of NESS" from the viewpoint of the Lorentz transformation of NESSs in Section \ref{sec:trfofTeff}.

In the above-mentioned example \cite{Gubser:2006nz,Nakamura:2013yqa,Hoshino:2014nfa}, the effective temperature at finite velocity is lower than the temperature of the heat bath.
However, we find that the proper effective temperature is {\it higher} than the proper temperature of the heat bath, as we show in detail in Section \ref{sec:prop_eff_temp}.
In fact, this is the case for a wide range of holographic models.
In this work, we use the natural unit $\hbar = c = k_B=1$.

\section{Proper temperature and Lorentz transformation of temperature in equilibrium systems}
Let us revisit the relation between the proper temperature and the Lorentz transformation of temperature in terms of holography.
We consider a system of gauge particles at equilibrium and at finite temperature.
The gauge theory is a strongly-coupled large-$N_c$ gauge theory in $d$ dimensions. 
The coordinates of the $d$-dimensional spacetime are $(t, \vec x)$, where $t$ is the time, and $\vec x=(x^1,\cdots,x^{d-1})$ denotes the space-like coordinates.
To simplify the notation, we refer to $x^1$ as $x$.

The gravity dual of the system is realized as a $(d+1)$-dimensional black-hole geometry \cite{Witten:1998qj} when the system is in the deconfinement phase.
The coordinates of the $(d+1)$-dimensional spacetime are $(t,\vec x, u)$, where $u$ is the radial direction.
The metric is given by
\eq{
	ds^2
&=
	g_{tt}(u)
	dt^2
	+
	g_{xx}(u)
	dx^2
	+
	g_{uu}(u)
	du^2
\\
&
	+
	\sum_{\mu \not = t,x,u}
	g_{\mu \mu}(u)
	(dx^\mu)^2
.
\label{eq:bgmetric}
}
In the vicinity of the horizon $u=u_{H}$, $g_{tt}$ and $g_{uu}$ are expanded as $
	g_{tt}
=
	-
	a
	(u-u_H)
	+
	O
	((u-u_H)^2)
$ and $
	g_{uu}
=
	{b}/\roundLR{u-u_H}
	+
	O
	(1)
$, respectively,
and the other components are positive and finite.
In our convention, $g_{tt}<0$ and $g_{uu}>0$ outside the horizon.
The Hawking temperature $T_0$ is obtained from the distribution of the Hawking radiation at the horizon\footnote{
The $u$-coordinate corresponds to the energy scale of the quantum field theory in the context of holography: the infrared behavior of the system is characterized by the geometry in the vicinity of the horizon.
}, $\exp \squareN{ -4 \pi \sqrt{b/a} \, \omega}$, where $\omega$ is the energy of the radiation.
It is given by
\eq{
	T_0
&=
	\frac{1}{4\pi}
	\sqrt{
	\frac{a}{b}
	}
.
\label{eq:T}
}
The Hawking temperature $T_0$ is identified with the temperature of the gauge theory.
The coordinate system $(t, \vec x)$ describes the frame of the observer $O_0$, who measures $T_0$ as the temperature.
In this frame, the probability distribution of the Hawking radiation is given by $\exp \squareLR{ -{E}_{0i}/T_0}$, where ${E}_{0i}$ is the energy of the $i$-th microscopic state.
The distribution function is isotropic, and hence we call this frame the rest frame.
The temperature $T_0$ is interpreted as the proper temperature (van Kampen's temperature).

Now, we consider an observer $O$ moving in the $x$-direction at constant velocity $\beta$ relative to the rest frame.
Let the observer $O$ measure $T$ as the temperature.
To obtain the temperature $T$, we perform a boost transformation 
$
	{dx\,'}^\mu
=
	\Lambda^\mu_{~\nu}
	dx^\nu
$, where
\eq{
	\Lambda^t_{~t}
&=
	\Lambda^x_{~x}
=
	\gamma(\beta)
,
\quad
	\Lambda^x_{~t}
=
	\Lambda^t_{~x}
=
	-
	\gamma(\beta)\beta
,
\\
	\gamma(\beta)
&=
	1/\sqrt{1-\beta^2}
.
\label{eq:Lambda}
}

The line element in the boosted frame is given by
\eq{
	ds^2
&=
	g'_{tt}
	dt'^2
	+
	2
	g'_{tx}
	dt'
	dx'
	+
	g'_{xx}
	dx'^2
\\
&
	+
	g_{uu}
	du^2
	+
	\sum_{\mu \not = t,x,u}
	g_{\mu \mu}
	(dx^\mu)^2
,
\label{eq:metricwithflow}
}
which is rewritten as
\eq{
	ds^2
&=
	\roundLR{
	g'_{tt}
	-
	\frac{
	g'^2_{tx}
	}{
	g'_{xx}
	}
	}
	dt'^2
	+
	g'_{xx}
	\roundLR{
	dx'
	+
	\frac{g'_{tx}}{g'_{xx}}
	dt'
	}^2
\\
&
	+
	g_{uu}
	du^2
	+
	\sum_{\mu \not = t,x,u}
	g_{\mu \mu}
	(dx^\mu)^2
,
\label{eq:diagofg}
}
in the vicinity of the horizon.
The coefficient of $dt'^2$ in (\ref{eq:diagofg}) is expanded as
\eq{
	g'_{tt}
	-
	\frac{
	g'^2_{tx}
	}{
	g'_{xx}
	}
&=
	-
	a'
	(u-u_H)
	+
	O((u-u_H)^2)
,
}
where $a' = a /\gamma^2$.
Hence the Hawking temperature in the boosted frame is given by
\eq{
	T
&=
	\frac{1}{4\pi}
	\sqrt{
	\frac{a'}{b}
	}
=
	\frac{T_0}{\gamma(\beta)}
.
\label{eq:relabetTandTprime}
}
The equation (\ref{eq:relabetTandTprime}) shows that the moving observer $O$ measures the lower temperature $T$ than $T_0$.
The transformation (\ref{eq:relabetTandTprime}) agrees with the Einstein-Planck's temperature, and hence this is a holographic derivation of the relation (\ref{eq:defofEinsteinsTprime}).
We can generalize the result (\ref{eq:relabetTandTprime}) by replacing $\beta$ with $\vec{\beta}$ in a general direction, since the system is rotational invariant at the rest frame.

From the viewpoint of the observer $O$, the probability distribution of the Hawking radiation (see Appendix \ref{sec:Tinboostedframe}) is given by
\eq{
	\exp
	\roundLR{
	-
	\frac{E_{i}-\vec{v} \cdot \vec{k}_{i}}{T}
	}
,
\label{eq:spectruminequilibrium}
}
where $E_{i}$ and $\vec{k}_{i}$ are the energy and the momentum of the $i$-th microscopic state, and $\vec{v} = -\vec{\beta}$ is the relative velocity of the system to the rest frame.
The transformation (\ref{eq:relabetTandTprime}) shows that $\gamma(v) T$ is equivalent to $T_0$ and hence is Lorentz invariant: $\gamma(v) T=\gamma(v) T_0/\gamma(\beta)=T_0$ since $\gamma(v)=\gamma(\beta)$.
In addition, we know $\gamma(v)(E_i- \vec{v}\cdot \vec{k}_i)$ is also Lorentz invariant.
Hence (\ref{eq:spectruminequilibrium}) is Lorentz invariant.
The probability distribution should be scalar, and hence we should employ (\ref{eq:spectruminequilibrium}) rather than $\exp \squareLR{ -{E}_i/{T}}$ in the relativistic theory.
This distribution is consistent with the relativistic distribution proposed in the previous works \cite{Juttner:1911,Cubero:2007}.

We emphasize that the rest frame, where the distribution function is isotropic, exists for each thermal equilibrium system.
In general, the distribution (\ref{eq:spectruminequilibrium}) is characterized by $\vec{v}$ and is anisotropic.
However, we can always perform a boost transformation with $\vec{v}$ and go back to the rest frame.
Hence, we refer to $T_0$ as the proper temperature.
This is an analog of the rest mass $m_0$.
The proper temperature $T_0$ characterizes each system at equilibrium.

We can interpret the velocity $\vec{v}$ in analogy to the chemical potential of a finite density system.
If we compare (\ref{eq:spectruminequilibrium}) with the grand canonical distribution at finite chemical potential $\mu$, $\exp \squareN{-(E_{i}-\mu N_{i})/T }$ with $N_{i}$ being the number of the particles of $i$-th microscopic state, $\vec{v}$ can be understood as a ``chemical potential for momentum''.
In the relativistic thermodynamics proposed by Einstein and Planck \cite{Einstein:1907,Planck:1908}, the first law of thermodynamics is given by $dE=TdS-PdV+\vec{v}\cdot d\vec{k}$, where $S$, $P$, $V$, and $\vec{k}$ are the entropy, the pressure, the volume, and the momentum of the system, respectively (see Appendix \ref{sec:first_law} for more details).
Then our spectrum (\ref{eq:spectruminequilibrium}) suggests that the term $\vec{v} \cdot d\vec{k}$ in $dE=TdS-PdV+\vec{v}\cdot d\vec{k}$ is understood in analogy with the term $\mu dN$, where $N$ is the number of the particles in the system.

For later use, we consider a new observer $\tilde{O}$ moving at velocity $\tilde{\beta}$ in the $x$-direction 
relative to the observer $O$.
The additional boost transformation with $\tilde{\beta}$ on top of (\ref{eq:Lambda}) yields
\eq{
	\tilde{T} 
&=
	\frac{T}{(1-v \tilde{\beta})\gamma(\tilde{\beta}) }
,
\label{eq:Ttilde}
}
where $\tilde{T}$ is the temperature measured by the observer $\tilde{O}$.
If we replace $(v, \tilde{\beta})$ by $(0,\beta)$, (\ref{eq:Ttilde}) returns to (\ref{eq:relabetTandTprime}).
We can generalize (\ref{eq:Ttilde}) by replacing $v \tilde{\beta}$ with the inner product $\vec{v} \cdot \vec{\tilde{\beta}}$.
We obtain
\eq{
    \gamma(\tilde{v})\tilde{T} = T_0,
}
where $\tilde{v}$ is the speed of the system relative to $O_0$ and $\gamma(\tilde{v})= (1-\vec{\tilde \beta}\cdot \vec{v})\gamma(\tilde{\beta})\gamma(v)$.

\section{Proper effective temperature in nonequilibrium steady states}
\label{sec:trfofTeff}
From now on, we consider boost transformation of the effective temperature of nonequilibrium steady states (NESSs).
A NESS is a nonequilibrium state where the macroscopic variables do not evolve in time.
To realize a NESS, we need three ingredients: a heat bath, a system we study, and an external force. The external force drives the system out of equilibrium, but the heat produced in the system dissipates into the heat bath, so that the system is steady from the macroscopic point of view.
Even in a NESS, it is possible to define a ``temperature'' as the ratio between the fluctuation and the dissipation; it is referred to as effective temperature \cite{Cugliandolo:2011}.

Let us consider a NESS in a system at finite charge density $\rho$, where both positive and negative charge carriers are immersed in a heat bath.
We apply the external electric field $\mathcal{E}$ along the $x$-direction.
Let the constant current $J$ flow in the same direction.
For simplicity, we consider the boost transformation along the electric field in this work, so that the magnetic field is not induced.

This system can be described in holography \cite{Karch:2007pd}.
Let us describe the system from the viewpoint of the observer $O_0$.
Then, the heat bath is mapped to the black hole geometry (\ref{eq:bgmetric}).
The sector of the charge carriers is mapped to a D-brane, which is defined in the framework of superstring theory.
The effective action of the D-brane is described by using the induced metric $h_{ab}$ and the U(1) gauge field strength ${F}_{ab}$ (see Appendix \ref{sec:DBI} for details).

The effective temperature in the presence of constant current is studied by using holography \cite{Kim:2011qh, Sonner:2012if, Nakamura:2013yqa, Hoshino:2014nfa}.
To obtain the effective temperature, we consider the linearized equations of motion for fluctuations on the D-brane.
For example, if we consider a scalar fluctuation, the linearized equation of motion is in the form of a Klein-Gordon equation on a curved geometry: we call this ``geometry" as the effective geometry.
The effective geometry differs from the spacetime geometry in general.
The metric of the effective geometry is given by the open-string metric, which is defined as $G_{ab}=h_{ab}-(2\pi\alpha^{\prime})^{2}(Fh^{-1}F)_{ab}$, in our setup.
One finds that the effective geometry has an effective horizon which plays a role of black-hole horizon for the fluctuations.
The location of the effective horizon differs from that of the horizon of the spacetime geometry when the electric field is turned on.
We consider the linearized equations in the vicinity of the effective horizon, $u=u_*$.  
Then, the effective temperature is given by the Hawking temperature, which is evaluated at the effective horizon\footnote{
The fluctuation-dissipation relation for equilibrium systems is obtained from the analysis of linearized fluctuations on the bulk geometry \cite{deBoer:2008gu}.
If we apply the same formalism for our fluctuations of NESS, we obtain the fluctuation-dissipation that is characterized by the effective temperature.
}.

The Hawking temperature is read from the open-string metric ${G}_{{a} {b}}$ in parallel with (\ref{eq:T}):
\eq{
	T_{*}
=
	\frac{1}{4\pi}
	\sqrt{\frac{a_*}{b_*}}
,
}
where $a_*$ and $b_*$ are given by $
	\mathcal{G}_{\hat{t} \hat{t}}
\equiv
	\roundN{{G}_{tt}{G}_{xx}-{G}_{xt}^2}/{{G}_{xx}}
=
	-
	a_*
	(u-u_*)
	+
	O
	((u-u_*)^2)
$
and
$
	\mathcal{G}_{\hat{u} \hat{u}}
\equiv
	{
	-\det {G}
	}/\squareN{
	{g}_{xx}^2
	\roundLR{
	{G}_{xt}^2
	-
	{G}_{tt} 
	{G}_{xx}
	}
	}
=
	b_*/(u-u_*)
	+
	O
	(1)
$ 
in the vicinity of $u_*$ (see Appendix \ref{sec:Teff} for details).
The effective temperature $T_{*}$ is measured by the observer $O_0$.

Next, let us consider the effective temperature $T_*^{\,\prime}$ measured by the moving observer $O$. We perform the boost transformation of the open-string metric $G_{ab}$ as
$
	G_{ a  b}^{\,\prime}
=
	\roundLR{
	\Lambda^{-1}
	}^{ c}_{~ a}
	\roundLR{
	\Lambda^{-1}
	}^{ d}_{~ b}
	G_{ c  d}
$,
where $\Lambda^{a}_{~ b}$ are given by (\ref{eq:Lambda}). 
Then we rearrange $G_{ a  b}^{\,\prime}$ (see (\ref{eq:diagonalizedminicoordinatecomponents}), (\ref{eq:defofQs}) and (\ref{eq:diagmetric}) in Appendix \ref{sec:Teff} for details).
One finds that the ``time-time" component of the metric $\mathcal{G}^{\prime}_{\hat{t} \hat{t}}$ is expanded as $\mathcal{G}^{\prime}_{\hat{t} \hat{t}}=-a_*^{\prime}(u-u_*)+O((u-u_*)^{2})$ in the vicinity of the effective horizon, where
$
	a_*'
=
	{
	a_*
	}/{
	\roundLR{
	1
	-
	v_m
	\beta
	}^2
	\gamma^2(\beta) 
	}
$ with $
	v_m 
=
	-
	{G}_{xt}/{G}_{xx}
	|_{u=u_*}
$. 
Then the effective temperature $T_*^{\,\prime}$ in the boosted frame is obtained as 
\eq{
	T^{\,\prime}_{*}
=
	\frac{1}{4\pi}\sqrt{\frac{a_*'}{b_*}}
&=
	\frac{
	T_*
	}{
	\roundLR{
	1
	-
	v_m
	\beta
	}
	\gamma (\beta)
	}
.
\label{eq:TstarprimeantTstar}
}
Note that (\ref{eq:TstarprimeantTstar}) has the same structure as (\ref{eq:Ttilde}).

We can define a proper effective temperature, a rest frame, and a velocity of NESS, in the same way as those of equilibrium. 
The distribution function of the fluctuations in the NESS was computed \cite{Hoshino:2017air} as
\eq{
	\exp \roundLR{ -\frac{E_{i}-v_m k_{i}}{T_*}}
,
\label{eq:probdensities}
}
where $E_{i}$ and $k_{i}$ are the energy and the $x$ component of the momentum of the $i$-th fluctuation in consideration, respectively.
If we perform a boost transformation with $\beta=v_{m}$,  the distribution function (\ref{eq:probdensities}) goes to 
\eq{
    \exp \roundLR{ - \frac{E_{*0i}}{{T_{*}}_{0}}}
,
\label{eq:probdensitiesatrestofNESS}
}
where $E_{*0i}$ is the energy of the $i$-th state in the new frame.
We have defined 
\eq{
	T_{*0}
\equiv
	T_*^{\,\prime}
	|_{\beta=v_m}
=
	\gamma (v_m)
	T_*
,
\label{eq:properefftemp}
}
which is a boost invariant quantity of the NESS.
Hence $T_{*0}$ can be interpreted as the ``proper effective temperature" in parallel with the proper temperature $T_0$. 
We call this frame the ``rest frame of the NESS," since the distribution function is isotropic.
Compared to (\ref{eq:spectruminequilibrium}), the distribution function (\ref{eq:probdensities}) states that the system has a relative velocity $v_{m}$ with respect to the rest frame of the NESS.
This is consistent with the interpretation of $v_{m}$ from the phenomenological viewpoint \cite{Hoshino:2017air}, where  $v_m$ is referred to as the mean velocity of NESS and is defined as the average of the velocities of the charge carriers.
Regarding $v_m$ as the relative velocity with respect to the rest frame of the NESS,
we understand that (\ref{eq:TstarprimeantTstar}) is the counterpart of (\ref{eq:Ttilde}).

\section{Comparison of proper effective temperature with proper heat-bath temperature}
\label{sec:prop_eff_temp}
Let us compare the proper effective temperature of a dragged object with the proper temperature of the heat bath for possible holographic setups. 

The superstring theory offers the near-horizon limit of the D$p$-brane geometry as a dual geometry of holography. For D$p$-branes, $p$ is a non-negative integer less than ten, and $p$ is even (odd) for the Type IIA  (IIB) superstring theory.
We consider $p\ge 1$ cases for our purpose
since $p$ is the number of the spatial directions of the heat bath. 
 Furthermore, it is appropriate to consider only for $p\le  4$, since the specific heat of the geometry is ill-defined or negative for $p\ge  5$ as we shall see below.

The near-horizon limit of the D$p$-brane geometry at finite temperature is given by~\cite{Itzhaki:1998dd} 
\begin{eqnarray}
ds^{2}=u^{(7-p)/2}\left[-\left(1-\left(\frac{u_{H}}{u}\right)^{7-p}\right)dt^{2}+d\vec{x}^{2}\right]+\frac{du^{2}}{u^{(7-p)/2}\left(1-\left(\frac{u_{H}}{u}\right)^{7-p}\right)}+u^{(p-3)/2}d\Omega^{2}_{8-p}.\:\:
\label{Dp-geometry}
\end{eqnarray}
The geometry has a compact sub-manifold $S^{8-p}$. The line element of the unit $S^{8-p}$ is given by $d\Omega^{2}_{8-p}$ in (\ref{Dp-geometry}). 
The Hawking temperature of the geometry is $T_0=\frac{7-p}{4\pi}u_{H}^{(5-p)/2}$.  
Let us consider the specific heat of the geometry. 
The entropy of the geometry is proportional to $u_{H}^{(7-p)p/4}=\left(\frac{4\pi}{7-p}T_0\right)^{(7-p)p/2(5-p)}$. One finds that the specific heat of the geometry is well-defined and positive only for $p\le 4$.
Therefore, let us focus on the $1\le p\le 4$ cases. 

The superstring theory offers a fundamental string (F1 string) or a D-brane as a natural candidate for the gravity dual of the dragged object. 
For D-branes, we consider D$(q+1+n)$-brane where $q$ denotes the spatial dimensionality of the dragged object. Then $q$ must satisfy $0\le q\le p-1$ since we need at least one direction along which we drag the object. Taking account of the above-mentioned constraint for $p$, we find $0\le q\le p-1 \le 3$. 
The D$(q+1+n)$-brane wraps the $S^{n}$ part of the $S^{8-p}$, thus $n\le 8-p$. Of course, $q+1+n\le 9$, and hence $q\le 8-n$. The configuration of the $S^{n}$ part of the D$(q+1+n)$-brane is determined by the equation of motion, hence it is non-trivial in general. However, we consider only for the cases where the D$(q+1+n)$-brane wraps the equatorial part of the $S^{8-p}$ for simplicity.

Let us examine the relation between the proper effective temperature of the dragged object and the proper temperature of the heat bath for possible holographic setups within the above constraints.

\subsection{F1 string}

When the gravity dual of the dragged object is an F1 string, the effective temperature is given by~\cite{Nakamura:2013yqa} as
\begin{eqnarray}
T_{*}=\left(1-v^{2}\right)^{1/(7-p)}T_0\le T_0,
\end{eqnarray}
where $v$ is the velocity of the dragged object with respect to the rest frame of the heat bath.
However, the proper effective temperature $T_{*0}$ is found to be
\begin{eqnarray}
T_{*0}=T_{*}\left(1-v^{2}\right)^{-1/2}=\left(1-v^{2}\right)^{(p-5)/2(7-p)}T_0\ge T_0,
\end{eqnarray} 
since we consider $p\le 4$. The equality holds only for $v=0$.

\subsection{D$(q+1+n)$-brane}

For D$(q+1+n)$-branes, the effective temperature is given by
\begin{eqnarray}
T_{*}=\left(1-v^{2}\right)^{1/(7-p)}\left(1+C v^{2}\right)^{1/2}T_0,
\end{eqnarray}
where $C=\frac{1}{2}(q+3-p+n\frac{p-3}{7-p})$~\cite{Nakamura:2013yqa}.
The proper effective temperature is then given by
\begin{eqnarray}
T_{*0}=\left(1-v^{2}\right)^{(p-5)/2(7-p)}\left(1+C v^{2}\right)^{1/2}T_0.
\end{eqnarray}
We have already seen that $\left(1-v^{2}\right)^{(p-5)/2(7-p)}\ge 1$ within our constraint. Therefore, $T_{*0}\ge T_0$ holds as far as $C\ge 0$. Let us examine the case where $C<0$.

\subsubsection{$p=1$ case}

In this case, only $q=0$ is possible, and $2C=2-n/3$. If $n\ge 7$, $C$ may be negative. Taking account of the constraint $n\le 8-p=7$, only the possible chance for negative $C$ is the case of $n=7$. However, this is impossible, since the D1-brane and the D$(0+1+7)$-brane cannot coexist in the framework of the superstring theory: the D$p$-brane of even number of $p$ exists in Type IIA superstring theory whereas the D$p$-brane of odd number of $p$ exists in Type IIB superstring theory. 

\subsubsection{$p=2$ case}

In this case, $q=0$ and $q=1$ are possible. We have $2C=(q+1-n/5)$ and $n\le 8-p=6$. For $q=0$, $2C=(1-n/5)$ which can be negative only for $n=6$. However, the D$2$-brane and D$(0+1+6)$-brane cannot coexist.
For $q=1$, $2C=(2-n/5)>0$ since $n\le 6$.

\subsubsection{$p=3$ case}

In this case, $2C=q$ which is always non-negative.

\subsubsection{$p=4$ case}

In this case, $2C=q-1+n/3$, which can be negative only when $q=0$ and $n$ is for $n=0, 1, 2$. For $q=0$, the possible D$(1+n)$-brane which may make $C$ negative and can coexist with the D4-brane is the D2-brane where $n=1$. For this case, we have $C=-1/3$ and the proper effective temperature is given by
\begin{eqnarray}
T_{*0}=\left(1-v^{2}\right)^{-1/6}\left(1-\frac{1}{3} v^{2}\right)^{1/2}T_0.
\end{eqnarray}
One finds that $T_{*0}\ge T_0$ since $v^{2}\le 1$.

\vspace{4ex}
Therefore, we find that $T_{*0}\ge T_0$ for all the possible cases we consider within our assumptions.

\section{Conclusions and Discussions}
\label{sec:disscussion}
Using holography, we have derived the boost transformation of the temperature (\ref{eq:relabetTandTprime}) and (\ref{eq:Ttilde}), and that of the effective temperature (\ref{eq:TstarprimeantTstar}) without relying on thermodynamics.
For systems at equilibrium, the transformation of temperature (\ref{eq:relabetTandTprime}) agrees with that of the Einstein-Planck's temperature,
and $\gamma(v) T$ is equivalent to the proper temperature (van Kampen's temperature) $T_0$.
We have shown that the Lorentz invariant distribution (\ref{eq:spectruminequilibrium}) is consistent with the Einstein-Planck's first law of thermodynamics given in \cite{Einstein:1907,Planck:1908}:
the velocity $ \vec{v}$ of the system at equilibrium can be interpreted as the ``chemical potential for momentum" in analogy with the grand canonical ensemble.

In NESSs, the boost transformation of effective temperature (\ref{eq:TstarprimeantTstar}) is essentially the same as that of temperature (\ref{eq:Ttilde}) in equilibrium.
From the distribution function (\ref{eq:probdensities}), we have interpreted $v_m$ as the mean velocity of the NESS, that agrees with the mean velocity obtained phenomenologically in \cite{Hoshino:2017air}.
We have defined the rest frame of NESS as the frame where the mean velocity is zero.
The rest frame of NESS is different from that of the heat bath in general.
In addition, we have defined the proper effective temperature $T_{*0}$.
The product $\gamma(v_m)T_*$ is equivalent to the proper effective temperature $T_{*0}$ and hence is boost invariant. 

We have examined the relation between the proper effective temperature $T_{*0}$ of the dragged object and the proper temperature $T_0$ of the heat bath, in the interesting examples discussed in \cite{Nakamura:2013yqa} where $T_*$ can be lower than $T_0$.
We have found that $T_{*0}\ge T_0$ for all the models considered in ref.~\cite{Nakamura:2013yqa}, as far as the heat capacity of the heat bath is positive ($p<5$) and the combinations of $p$, $q$ and $n$ are consistent with the superstring theory.
This suggests that the interactions between the dragged object and the heat bath contribute to {\em raising} the proper effective temperature $T_{*0}$ from $T_0$.

We have two remarks.
First, let us consider a four-vector $\beta_T^\mu $ discussed in \cite{Kampen:1968}, which is defined by
$
	\beta_T^\mu 
= 
	u^\mu /	T_0
= 
	(1,\vec{v})/ T
$,
where $u^\mu$ is the four-velocity of the heat bath given by $u^\mu = \gamma(v) (1,\vec{v})$.
Since $T_0$ is Lorentz invariant, $\beta_T^\mu$ is well defined as a four-vector. 
One finds the following relationship holds:
$
	\beta_{T0}^2
=
	-
	\eta_{\mu\nu}
	\beta_T^\mu
	\beta_T^\nu
$,
where $\beta_{T0}=1/T_{0}$ and $\eta_{\mu\nu}=\diag (-1,1,\cdots,1)$.
Using $\beta_T^\mu$,
we can rewrite (\ref{eq:spectruminequilibrium}) as $\exp (\eta_{\mu\nu} \beta_T^\mu k^\nu_i)$, where $k_i^\nu = (E_i,\vec{k}_i)$.
In NESSs, let us define 
\eq{
	\beta_{T*}^\mu
=
	\frac{
	u_m^\mu
	}{
	T_{*0}
	}
,
}
where $u_m^\mu = \gamma(v_m) (1,{v}_m,0,\cdots,0)$.
We obtain the relation $
	\beta_{T*0}^2
=
	-
	\eta_{\mu\nu}
	\beta_{T*}^\mu
	\beta_{T*}^\nu
$, where $\beta_{T*0}=1/T_{*0}$, and (\ref{eq:probdensities}) is rewritten as $\exp (\eta_{\mu\nu} \beta_{T*}^\mu k^\nu_i)$.

Secondly, let us discuss the fluctuation-dissipation relation (FDR) in the NESSs.
When the mean velocity is zero, the FDR is given by $G_{ij}^{\text{sym}} (\omega) = - \coth (\omega /2T_*) \im G_{ij}^{\text{R}} (\omega)$, where $G_{ij}^{\text{sym}} (\omega)$ is the symmetrized Wightman function given by $\int dt e^{-i\omega t} \angleN{ \curlyN{\delta \mathcal{O}_i(t), \delta\mathcal{O}_j(0)}}/2 $ for a set of operators $O_i$ \cite{Nakamura:2013yqa,Gursoy:2010aa,Sonner:2012if}.
$\delta\mathcal{O}_j(t)$ is defined as $ \mathcal{O}_j(t)-\angleN{\mathcal{O}_j(t)}$.
$G_{ij}^{\text{R}} (\omega)$ is the retarded Green's function given by $-i\theta(t) \int dt e^{-i\omega t} \angleN{ \squareN{ \delta \mathcal{O}_i(t) , \delta\mathcal{O}_j(0) } } $.
In our case where the mean velocity is finite, the FDR should be a covariant form as
\eq{
	G_{ij}^{\text{sym}} (k) 
&= 
	- 
	\coth \roundLR{
	-
	\frac{
	\eta_{\mu\nu} 
	\beta_{T*}^\mu 
	k^\nu
	}{
	2
	}
	} 
	\im G_{ij}^{\text{R}} (k)
.
}

When the system is relativistic, the effect of the boost transformation of effective temperature is not negligible, and the proper effective temperature plays a crucial role in NESSs.

\section*{Acknowledgments}
We thank S. Kinoshita for fruitful discussions.
We also thank S. Aoki, K. Murata, and H. Okamoto for valuable comments.
The research of H. H. is supported by the ISIRD grant 9-289/2017/IITRPR/704.
The work of S. N. was supported in part by JSPS KAKENHI Grant Number JP16H00810, JP19K03659, JP19H05821, and the Chuo University Personal Research Grant.

\appendix
\renewcommand{\thesection}{\Alph{section}}
\renewcommand{\thesubsection}{\Alph{section}.\arabic{subsection}}
\renewcommand{\theequation}{\Alph{section}.\arabic{equation}}
\section{Review of three relativistic thermodynamics at equilibrium}
\label{sec:first_law}

We review the three formulations of relativistic thermodynamics for a system at equilibrium, along the lines of \cite{Dunkel:2009tla,Farias:2017}.
They are proposed by Einstein and Planck \cite{Einstein:1907,Planck:1908}, by Landsberg and van~Kampen \cite{Landsberg:1966, Kampen:1968}, and by Ott \cite{Ott:1963}.
In each formulation, we consider a homogeneous system of finite volume at equilibrium.
We introduce two frames:
let a frame $\Sigma_0$ be the rest frame of the equilibrium system; let a frame $\Sigma$ be moving in the $x$-direction at constant velocity $-\vec{v}$ relative to $\Sigma_0$.
The gamma factor is given by $\gamma \equiv 1/\sqrt{1-v^2}$.

Before going into the relativistic definitions, let us note the first law of ordinary thermodynamics in the rest frame $\Sigma_0$:
\eq{
    \delta
    Q_0
&=
    T_0
    dS_0
=
    dE_0
    +
    P_0
    dV_0
,
\label{eq:firstlawatsigmazero}
}
where $\delta Q_0,~T_0,~S_0,~E_0,~P_0$, and $V_0$ are the heat, the temperature, the entropy, the energy, the pressure, and the volume of the system in the rest frame $\Sigma_0$, respectively\footnote{
The heat $\delta Q_0$ is the energy transfer in thermodynamic processes between one equilibrium state and another equilibrium state.}.
These quantities are observed on a hyperplane $\mathcal{T}_0$ where the time $t_0$ in the frame $\Sigma_0$ is constant\footnote{The hyperplane $\mathcal{T}_0$ corresponds to $\mathfrak{T}(\xi^0{})$ in \cite{Dunkel:2009tla}.}.
The question is how the formulation (\ref{eq:firstlawatsigmazero}) is transformed when we change the frame from $\Sigma_0$ to $\Sigma$.
We review that three different Lorentz transformations of temperature originate from different definitions of ``heat" and temperature in $\Sigma$ as below.
Note that, however, the three formalisms are translatable into each other as shown in Appendix \ref{sec:local}.

\subsection{Einstein-Planck's thermodynamics}
Einstein \cite{Einstein:1907} and Planck \cite{Planck:1908} introduce another hyperplane $\mathcal{T}$, and they define thermodynamic quantities in the frame $\Sigma$ on $\mathcal{T}$.
The hyperplane $\mathcal{T}$ is defined as $t =  \text{constant}$, where $t$ denotes the time in the frame $\Sigma$.

They define the heat $\delta Q$ in the frame $\Sigma$ as
\eq{
	\delta Q
&\equiv
	dE
	-
	v
	dk
	+
	P
	dV
,
\label{eq:defofheatEinstein}
}
where $P$ is the pressure in the frame $\Sigma$.
$E$, $k$, and $V$ are the energy, the momentum, and the volume evaluated on $\mathcal{T}$ in the frame $\Sigma$.
They are given by (see also (\ref{eq:VEk}))
\eq{
	P&=P_0
,
\\
	E
&=
	\gamma
	\roundLR{
	E_0
	+
	v^2
	P_0
	V_0
	}
,
\\
	k
&=
	\gamma
	v
	\roundLR{
	E_0
	+
	P_0
	V_0
	}	
,
\\
	V &= V_0/\gamma
,
\label{eq:defofhyperplaneEinstein}
}
where $P$ is treated as a Lorentz scalar \cite{Einstein:1907,Planck:1908}.
Note that $(E,~\vec k)$ is not a Lorentz transform of $(E_0,~ \vec 0)$, since they are defined on the different hyperplanes $\mathcal{T}$ and $\mathcal{T}_0$ (see also Appendix \ref{sec:local}).

The temperature $T$ in the frame $\Sigma$ is defined as 
\eq{
	T
	dS
&\equiv
	\delta Q
,
\label{eq:defofTEinstein}
}
where $S$ is the entropy in the frame $\Sigma$.
We assume that the entropy is invariant,
\eq{
	S = S_0
,
}
regardless of the choice of the hyperplane, since the entropy is given by the number of states of the entire system.

The Einstein-Planck's Lorentz transformation of $T$ is obtained as follows.
Using (\ref{eq:defofhyperplaneEinstein}), we have
$
	dE
	- 
	v 
	dk
=
	\gamma^{-2}
	P_0
	V_0
	d\gamma
	+
	\gamma^{-1}
	d
	E_0
$.
Then the definition of heat (\ref{eq:defofheatEinstein}) yields
\eq{
	\delta 
	Q
&=
	dE
	-
	v
	dk
	+
	P
	dV
\\
&=
	\gamma^{-1}
	\roundLR{
	dE_0
	+
	P_0
	dV_0
	}
=
	\gamma^{-1}
	\delta Q_0
.
}
Putting this together with the definition of temperature (\ref{eq:defofTEinstein}), we obtain
\eq{
	T
=
	\frac{T_0}{\gamma}
.
\label{eq:defofEinsteinsTprime}
}
This is the Einstein-Planck's temperature.

\subsection{Van Kampen's thermodynamics}
We revisit the first law of thermodynamics and temperature originally proposed by Landsberg  \cite{Landsberg:1966} and van~Kampen \cite{Kampen:1968} (see also the review \cite{Dunkel:2009tla}).
In this thermodynamics, all physical quantities are defined on the common hyperplane $\mathcal{T}_0$.

Following van~Kampen \cite{Kampen:1968}, we define thermal energy-momentum transfers $\delta Q_0^\mu$ and $\delta Q^\mu$ as
\eq{
	\delta Q_0^\mu
&\equiv
	dk_0^\mu
	-
	\delta A_0^\mu
,
\qquad
	\delta Q^\mu
\equiv
	dk^\mu
	-
	\delta A^\mu
,
\label{eq:defofTEMTfvector}
}
where
\eq{
	d k_0^\mu
&\equiv
	(dE_0,0)
,
\qquad
	dk^\mu
\equiv
	\gamma (dE_0, v dE_0)
=
	dE_0 
	\cdot
	u^\mu
,
\qquad
	u^\mu
=
	\gamma (1, v,0,0 )
,
\\
	dA_0^\mu
&\equiv
	-
	(P_0dV_0,0)
,
\qquad
	dA^\mu
\equiv
	-
	\gamma(P_0dV_0,v P_0dV_0)
=
	-
	P_0dV_0 
	\cdot
	u^\mu
.
\label{eq:defofdkanddA}
}
All these quantities are defined on the hyperplane $\mathcal{T}_0$.
Hence $dk_0^\mu$ and $ dA_0^\mu$ transform into $dk^\mu$ and $ dA^\mu$ as four-vectors under the Lorentz transformation, respectively.
These imply that $\delta Q^\mu$ is also a four-vector.
We employ (\ref{eq:defofTEMTfvector}) instead of (\ref{eq:defofheatEinstein}) in the present formulation.

In the frame $\Sigma$, temperature $T$ is defined by $-u_\mu \delta Q^\mu$ instead of (\ref{eq:defofTEinstein}):
\eq{
	TdS
&\equiv
	-
	u_\mu 
	\delta Q^\mu
=
	-
	u_{0\mu }
	\delta Q_0^\mu
=
	T_0dS_0
,
}
where $	u_0^\mu = (1, 0,0,0 )$, and $S$ is the entropy defined in the frame $\Sigma$ on the hypersurface $\mathcal{T}_0$.
The entropy is invariant,
\eq{
	S = S_0
,
}
under the Lorentz transformation, since the entropy is given by the number of states.

Hence the van Kampen's Lorentz transformation of temperature is given by
\eq{
	T&=T_0
.
\label{eq:defofTpbyKampen}
}

\subsection{Ott's thermodynamics}
We revisit the first law of thermodynamics and temperature proposed by Ott \cite{Ott:1963} (see also the review \cite{Dunkel:2009tla}).
As well as the previous case, we consider only the 
hyperplane $\mathcal{T}_0$.
We also employ the definitions (\ref{eq:defofTEMTfvector}) and (\ref{eq:defofdkanddA}).
The difference between the Ott's Lorentz transformation and the van Kampen's one comes from just the difference of the definition of temperature.

In the frame $\Sigma$, the Ott's temperature $T$ is defined by the time-component of $\delta Q^\mu$:
\eq{
	TdS
&\equiv
	\delta Q^t
=
	\gamma
	\delta Q_0^t
=
	\gamma
	T_0
	dS_0
.
}
Hence the Ott's Lorentz transformation of temperature is given by
\eq{
	T
&=
	\gamma
	T_0
.
\label{eq:defofTpbyOtt}
}

\section{Relationship between the three relativistic thermodynamics of finite volume system and the relativistic thermodynamics at local equilibrium}
\label{sec:local}
Here we reproduce the three descriptions of thermodynamics given in Appendix \ref{sec:first_law}, the relativistic thermodynamics of perfect fluid at local equilibrium.
In this formalism, thermodynamic quantities are defined in the local rest frame of the fluid element \cite{Israel:1976tn}.
A perfect fluid is described by the energy-momentum tensor $T^{\mu\nu}$, a conserved particle flux $n^\mu$, and an entropy flux $s^\mu$:
\eq{
	T^{\mu\nu}
&=
	\epsilon_0
	u^\mu
	u^\nu
	+
	P_0
	(
	u^\mu
	u^\nu
	+
	\eta^{\mu\nu}
	)
,
\\
    \partial_\mu 
    n^\mu
&=
    0
,
\qquad
    \partial_\mu 
    s^\mu
=
    0
,
\label{eq:emtensor}
}
where $\epsilon_0$ and $P_0$ are the energy density and the pressure in the local rest frame of the fluid element, $u^\mu$ is the four-velocity of the fluid element, and $\eta^{\mu\nu}$ is $\diag (-1,1,1,1)$. 

The thermodynamics at local equilibrium is given by
\eq{
	T_0
	s_0
&=	
	\epsilon_0
	+
	P_0
	-
	\mu_0
	n_0
,
\qquad
	T_0
	ds_0
=
	d\epsilon_0 
	-
	\mu_0
	d n_0
,
\label{eq:hdTD1}
}
where $T_0$, $s_0$, $\mu_0$, and $n_0$ are the proper temperature, the entropy density, the chemical potential, and particle number density, respectively.
Using a four-velocity of the fluid element $u^\mu$, we can rewrite (\ref{eq:hdTD1}) as
\eq{
	T_0
	s^\mu
&=
	\epsilon^\mu
	+
	P_0
	u^\mu
	-
	\mu_0
	n^\mu
,
\qquad
	T_0
	ds^\mu
=
	d\epsilon^\mu
	+
	P_0
	du^\mu
	-
	\mu_0
	d n^\mu
,
\label{eq:hdTD2}
}
where the flux of entropy $s^\mu$, of energy $\epsilon^\mu$, and of particle $n^\mu$ are written as
\eq{
	s^\mu
&=
	s_0 u^\mu
,
\qquad
	\epsilon^\mu
=
	\epsilon_0 
	u^\mu
,
\qquad
	n^\mu
=
	n_0 
	u^\mu
.
}

Employing them, we will reproduce the three formalisms of thermodynamics for a uniform and finite volume system at $\mu_0=0$: the Einstein-Planck's thermodynamics, the van Kampen's thermodynamics, and the Ott's thermodynamics discussed in Appendix \ref{sec:first_law}.
As we will see below, the difference of the three temperatures in Appendix \ref{sec:first_law} originates from merely their different definitions.

\subsection{Einstein-Planck's thermodynamics}
Here we reproduce the Einstein-Planck's thermodynamics.
We assume that the fluid is uniform.
As in Appendix \ref{sec:first_law}, we consider the two frames $\Sigma_0$ and $\Sigma$:
$\Sigma_0$ is the local rest frame of the uniform fluid;
$\Sigma$ denotes the frame moving at constant velocity $-\vec v$ relative to $\Sigma_0$. 
We also consider the two hyperplanes $\mathcal{T}_0$ and $\mathcal{T}$.

In the frame $\Sigma$, we define the volume $V$, energy $E$, and momentum $k$ evaluated on the hyperplane $\mathcal{T}$ as  
\eq{
	V
&=
	V_0/\gamma
,
\\
	E
&\equiv
	-
	\int_{\mathcal{T}(\xi{})}
	d\Sigma_\nu
	T^{t \nu}
=
	-
	\int_{\mathcal{T}(\xi{})}
	d\Sigma_t
	T^{t t}
=
	\gamma
	\roundLR{
	E_0
	+
	v^2
	P_0
	V_0
	}
,
\\
	k
&\equiv
	-
	\int_{\mathcal{T}(\xi{})}
	d\Sigma_\nu
	T^{x \nu}
=
	-
	\int_{\mathcal{T}(\xi{})}
	d\Sigma_t
	T^{x t}
=
	\gamma
	v
	\roundLR{
	E_0
	+
	P_0
	V_0
	}	
,
\label{eq:VEk}
}
where 
\eq{
	d\Sigma_\mu
&\equiv
	\frac{1}{3!}
	{\epsilon}_{\mu \nu_1 \nu_2 \nu_3}
	dx^{\nu_1}
	\wedge
	dx^{\nu_2}
	\wedge
	dx^{\nu_3}
,
}
$
	\epsilon_{txyz}
\equiv
	-1
$,
$	E_0
\equiv
	\epsilon_0
	V_0
$,
and 
$
	u^\mu
=
	\gamma
	(1,v,0,0)
$.
(\ref{eq:VEk}) agrees with (\ref{eq:defofhyperplaneEinstein}).

Let us integrate the second equation in (\ref{eq:hdTD2}) on $\mathcal{T}$.
The left-hand side becomes
\eq{
	\int_{\mathcal{T}(\xi{})}
	d\Sigma_\mu
	T_0
	ds^\mu
&=
	\int_{\mathcal{T}(\xi{})}
	d\Sigma_t
	T_0
	d(\gamma s_0)
=
	V
	T_0
	d(\gamma s_0)
,
\label{eq:left}
}
and the right-hand side becomes
\eq{
	\int_{\mathcal{T}(\xi{})}
	d\Sigma_\mu
	\squareLR{
	d\epsilon^\mu
	+
	P_0
	du^\mu
	}
&=
	\int_{\mathcal{T}(\xi{})}
	d\Sigma_t
	\squareLR{
	d(\gamma \epsilon_0)
	+
	P_0
	d\gamma
	}
\\
&=
	V
	\squareLR{
	d(\gamma \epsilon_0)
	+
	P_0
	d\gamma
	}
.
\label{eq:right}
}
Using (\ref{eq:left}), (\ref{eq:right}) and (\ref{eq:hdTD1}), and defining $S_0 \equiv s_0 V_0 = s V \equiv S$, $s \equiv \gamma s_0$, $	T
\equiv
	{T_0}/{\gamma}
$ and $P=P_0$, we have
\eq{
	T
	dS
&=
	dE
	- 
	v 
	dk
	+
	P
	dV
,
}
where $E,k,V$ are defined by (\ref{eq:VEk}).
The Einstein-Planck's thermodynamics is reproduced.

\subsection{Van Kampen's thermodynamics}
We will reproduce the van Kampen's thermodynamics.
For both frame $\Sigma_0$ and $\Sigma$, we define the thermodynamics on the same hyperplane $\mathcal{T}_0$.
Multiplying (\ref{eq:hdTD2}) by $V_0$, we have
\eq{
	T_0
	S_0
	u^\mu
&=	
	E_0
	u^\mu
	+
	P_0
	V_0
	u^\mu
.
}
The variation of this yields
\eq{
	T_0
	dS_0
	\cdot
	u^\mu
&=	
	dE_0
	\cdot
	u^\mu
	+
	P_0
	d
	V_0
	\cdot
	u^\mu
.
\label{eq:TdSua}
}
This is the thermal energy-momentum transfer, (\ref{eq:defofTEMTfvector}), in van Kampen's formalism.
Only the four-velocity $u^\mu$ changes under the Lorentz transformation.
We define the temperature as the Lorentz invariant quantity: $T\equiv T_0$.
The van Kampen's thermodynamics is reproduced.

\subsection{Ott's thermodynamics}
We define the ``heat" as (\ref{eq:TdSua}) for both frames.
Let the temperature be defined as
$
	T
\equiv
	\gamma
	T_0
$.
Then the Ott's thermodynamics is reproduced.

\section{Hawking temperature in the boosted frame}
\label{sec:Tinboostedframe}
We derive the Hawking temperature measured by the observer $O$, who moves in the $x$-direction at velocity $\beta$ relative to the rest frame of the system.
From the viewpoint of the observer $O$, the metric $g'_{\mu\nu}$ is given by (\ref{eq:metricwithflow}) in the main text, which has the off-diagonal component $g'_{tx}$.
Let us consider a free massless scalar field $\phi(t,\vec x,u)$, which obeys the equation of motion,
\eq{
&
	\roundBB{
	g^{\prime tt}
	\nabla_{t'}^2
	+
	2
	g^{\prime tx}
	\nabla_{t'}
	\nabla_{x'}
	+
	g^{\prime xx}
	\nabla_{x'}^2
\\
&\qquad
	+
	g^{ uu}
	\nabla_{u}^2
	+
	\sum_{\mu \not = t,x,u}
	g^{\mu \mu}
	\nabla_{\mu}^2
	}
	\phi
=
	0
,
\label{eq:EOMofphi}
}
where $\nabla_{\mu'}$ ($\nabla_{\mu}$) are the covariant derivatives with respect to $g'_{\mu\nu}$ ($g_{\mu\nu}$), 
$	
	g^{\prime tt}
=
	(\Lambda_{~t}^{t})^2
	g^{tt}
	+
	(\Lambda_{~x}^{t})^2
	g^{xx}	
$,
$
	g^{\prime tx}
=
	\Lambda_{~t}^{t}
	\Lambda_{~t}^{x}
	g^{tt}
	+
	\Lambda_{~x}^{t}
	\Lambda_{~x}^{x}
	g^{xx}
$, and
$	
	g^{\prime xx}
=
	(\Lambda_{~t}^{x})^2
	g^{tt}
	+
	(\Lambda_{~x}^{x})^2
	g^{xx}	
$ with 
\eq{
	\Lambda^t_{~t}
&=
	\Lambda^x_{~x}
=
	\gamma(\beta)
,
\quad
	\Lambda^x_{~t}
=
	\Lambda^t_{~x}
=
	-
	\gamma(\beta)\beta
,
\\
	\gamma(\beta)
&=
	1/\sqrt{1-\beta^2}
.
}

In (\ref{eq:EOMofphi}), the terms including $\nabla_{t'}$ and $\nabla_{x'}$ can be rewritten as
\eq{
	\squareLR{
	g^{\prime tt}
	\nabla_{t'}^2
	+
	2
	g^{\prime tx}
	\nabla_{t'}
	\nabla_{x'}
	+
	g^{\prime xx}
	\nabla_{x'}^2
	}
	\phi
&=
	\squareLR{
	\frac{
	g'_{xx}
	}{
	g'_{tt} 
	g'_{xx}
	-
	g'^2_{tx}
	}
	\roundLR{
	\nabla_{t'} 
	-
	\dfrac{g'_{tx}}{g'_{xx}}
	\nabla_{x'} 
	}^2
	+
	\frac{1}{g^{\prime}_{xx}}
	\nabla_{x'}^2
	}
	\phi
}
since $g'^{\mu\nu}$ are components of the inverse matrix of the metric $g'_{\mu\nu}$.

We follow the approach given in ref.~\cite{Srinivasan:1998ty},
where the spectrum of the radiation is derived by considering a tunneling effect in the semi-classical approximation.
First, we consider the equation (\ref{eq:EOMofphi}) in the vicinity of the horizon, which is approximated as
\eq{
	\squareLR{
	-
	\frac{
	1
	}{
	a'(u-u_H)
	}
	\roundLR{
	\partial_{t'} 
	+
	v
	\partial_{x'} 
	}^2
	+
	\partial_u
	\roundLR{
	\frac{u-u_H}{b}
	\partial_u
	}
	+
	\roundLR{\text{other terms}}
	}
	\phi
&=
	0
,
\label{eq:EOMforphi1}
}
where $u_H$ is the location of the horizon, and $v=-\beta=-{g'_{tx}}/{g'_{xx}}|_{u=u_H}$.
The constant $a'$ and $b$ are given by $
	\roundN{
	g'_{tt} 
	g'_{xx}
	-
	g'^2_{tx}
	}
	/
	g'_{xx}
=
	-
	a'
	(u-u_H)
	+
	O
	((u-u_H)^2)
$ and $
	g_{uu}
=
	{b}/\roundLR{u-u_H}
	+
	O
	(1)
$, respectively.
We employ the tortoise coordinate $dU \equiv du/(u-u_H)$ and obtain
\eq{
	\squareLR{
	-
	\frac{
	1
	}{
	a'(u-u_H)
	}
	\roundLR{
	\partial_{t'} 
	+
	v
	\partial_{x'} 
	}^2
	+
	\frac{1}{b(u-u_H)}
	\partial_U^2
	+
	\roundLR{\text{other terms}}
	}
	\phi
&=
	0
.
\label{eq:EOMforphi}
}
The (other terms) are $O(1)$, and hence they do not affect the solution in the vicinity of the horizon.
Employing the WKB approximation, we obtain the distribution of the scalar field as
\eq{
	\exp \roundLR{ -\frac{E_i- {v} {k}_i}{T}}
.
\label{eq:distributioninboostedframe}
}
$k_i$ and $E_i$ are the momentum in the $x$-direction and the energy of the $i$-th microscopic state measured by the moving observer $O$.
The Hawking temperature $T$ for $O$ is given by $
	4\pi
	T
=	
	\sqrt{
	{a'}/{b}
	}
$.
The result (\ref{eq:distributioninboostedframe}) is generalized
by replacing the velocity $\beta$ in the $x$-direction with $\vec{\beta}$ in a general direction, because of the rotational symmetry in the equilibrium system at the rest frame.
Then, the distribution function is given by
\begin{eqnarray}
	\exp
	\roundLR{
	-
	\frac{E_{i}-\vec{v} \cdot \vec{k}_{i}}{T}
	}
,
\end{eqnarray}
where $\vec{v} = -\vec \beta$, and $\vec{k}_{i}$ is the momentum of the $i$-th microscopic state.

\section{Gravity dual of NESS}
\label{sec:DBI}
Here, we describe a homogeneous NESS, where the macroscopic quantities are independent of the coordinates $(t,\vec{x})$.
To realize the NESS, we need three ingredients: a heat bath, a system we study, and an external force.
Let us consider a system that has both positive and negative charge carriers immersed in the heat bath in the presence of a constant electric field.
Suppose that the dissipation balances with the inflow of energy given by the external field to realize the NESS.

Let us describe the NESS, using holography \cite{Karch:2007pd}.
We consider a ten-dimensional black-hole spacetime.
We assume that the spacetime is decomposed into the following subspaces:
a $({d}+1)$-dimensional spacetime $(t,x^1,\cdots,x^{{d}-1},u)$, where ${d} < 9$;
a $(9-{d})$-dimensional compact space $(x^{{d}+1},\cdots,x^{9})$.
Now we are in the frame of the observer $O_0$, and
the metric of the $(d+1)$-dimensional spacetime is given by (\ref{eq:bgmetric}) in the main text.
This black hole geometry is dual to the heat bath.

The sector of the charge carriers is described by a D($q+1+n$)-brane \cite{Ammon:2009jt,Hoshino:2017air}, where ${d} \geq q \geq 1$.
We employ the probe approximation.
Let $\zeta^a$ be the coordinates on the worldvolume of the brane.
We employ the static gauge, $\zeta^a = (t,u,\vec{\tilde{x}},\vec \Omega)$:
the worldvolume extends in $\vec{\tilde{x}} = (x^{1}, \cdots, x^{q})$ and the $u$-direction, and wraps an $n$-dimensional subspace $\vec \Omega$ of the $(9-{d})$-dimensional compact manifold, where $9-{d} \geq n$.
The Dirac-Born-Infeld action for the probe D($q+1+n$)-brane is given by
\eq{
    S_{D(q+1+n)}
&=
-
    T_{D(q+1+n)}
    \int d^{q+2+n} \zeta
\\
&\qquad
\times
    e^{-\phi}
    \sqrt{-\det
    \roundLR{
    h_{ab}
    +
    2\pi \alpha'
    {F}_{ab}
    }
    }
,
\label{eq:DBIaction}
}
where $T_{D(q+1+n)}$ is the tension of the brane.
The dilaton field $\phi$ depends only on $u$.
The constant $\alpha'$ is related to the 't Hooft coupling $\lambda$.
For example, $\lambda = 2N_c T_{Dp} (2\pi \alpha')^{-2}$
when the bulk geometry is the D$p$-brane background \cite{Itzhaki:1998dd}, where $N_c$ is the rank of the gauge group.
The induced metric $h_{ab}$ and the U(1) gauge field strength ${F}_{ab}$ on the brane are defined by
$
    h_{ab}
=
    \partial_a X^\mu
    \partial_b X^\nu
    g_{\mu \nu}
$ and $
    {F}_{ab}
=
    \partial_a {A}_{b}
    -
    \partial_b {A}_{a}
,
$
where $X^\mu(\zeta)$ represents the configuration of the D($q+1+n$)-brane, and $A_a(\zeta)$ is the U(1) gauge field.
We assume that the Wess-Zumino term is not switched on.

We assume that $F_{ab}$ and $X^\mu$ depend only on $u$.
For the gauge field, we take the following ansatz:
$A_t$ depends on only $u$;
$
	A_x(t,u) = -\mathcal{E} t +a_x(u)
$;
the other components of the gauge field are switched off.
The constant $\mathcal{E}$ is the external electric field acting on the charge carriers.

The Gubser-Klebanov-Polyakov-Witten prescription gives the following identifications \cite{Gubser:1998bc,Witten:1998qj,Karch:2007pd}:
\eq{
	\frac{\delta \mathcal{L}}{\delta {A}_t'}
&=
	\rho
,
\qquad
	\frac{\delta \mathcal{L}}{\delta {A}_x'}
=
	J
,
\label{eq:integratedEOM}
}
where the prime denotes $\partial/\partial u$, and $
	\mathcal{L}
=
	T_{D(q+1+n)}
	\int d^{n} \zeta
	e^{-\phi}
	\sqrt{-\det
	\roundLR{
	h_{ab}
	+
	2\pi \alpha'
	{F}_{ab}
	}
	}
$.
$\rho$ and $J$ are the expectation values of the charge density and the current density, respectively.
The reality condition for the gauge fields yields the relation between $J$ and $\mathcal{E}$,
\eq{
	J
&=
	\frac{2\pi \alpha'}{{h}_{xx}}
	\sqrt{
	{\rho}^2
	+
	\tilde{V}^2
	}
	~\mathcal{E}
	\Bigg|_{u=u_*}
,
\label{eq:JinD3D7withB3}
}
where $u_*$ is defined by
$
	\absLR{h_{tt}(u_*)}
	h_{xx}(u_*)
=
	(2\pi\alpha')^2
	\mathcal{E}^2
$ \cite{Karch:2007pd}.
Here, 
$
	\tilde{V}
=
	2\pi\alpha'
	T_{D(q+1+n)}
	\int d^n \zeta
	e^{-\phi}
	h_{xx}^{q/2}
	\sqrt{h_{\Omega}}
$, where $d^n\zeta \sqrt{h_{\Omega}}$ is the volume element of the worldvolume along the compact directions, which may depend on the brane configuration.
(See also ref. \cite{Hoshino:2017air}.)
This means that $\sqrt{h_{\Omega}}$, hence $\tilde{V}$, contains dynamical degrees of freedom, which are determined by the Euler-Lagrange equations.

\section{Derivation of effective temperature}
\label{sec:Teff}
Let us consider the open-string metric ${G}_{ab}$ discussed in Section \ref{sec:trfofTeff} in the main text.
We focus on the $3\times 3$ part of the open-string metric on the $(t,x,u)$ coordinates.
In our setup, $G_{ab}$ has off-diagonal components,
$G_{ut}=G_{tu} =-(2\pi\alpha^{\prime})^{2} F_{tx} h^{xx} F_{xu}$ and $G_{xt}=G_{tx} =-(2\pi\alpha^{\prime})^{2} F_{tu} h^{uu} F_{ux}$ where $F_{xt} = \mathcal{E}$, since $F_{ut}$ ($F_{ux}$) is switched on when $\rho \not = 0$ ($J\not = 0$).
On this geometry, the line element can be written as\footnote{
Note that the deformation given in (\ref{eq:diagonalizedminicoordinatecomponents}) is not a coordinate transformation to diagonalize the metric for arbitrary $u$.
This is because $dx+Q^{\hat{x}}_{~t}dt+Q^{\hat{x}}_{~u}du$ is not integrable since $Q^{\hat a}_{~ b}$ depends on $u$.
}
\eq{
	ds_{3 \times 3}^2
&=
	\mathcal{G}_{\hat{t} \hat{t}}
	\roundLR{
	dt
	+
	Q^{\hat{t}}_{~u}
	du 
	}^2
	+
	\mathcal{G}_{\hat{x} \hat{x}}
	\roundLR{
	dx 
	+ 
	Q^{\hat{x}}_{~t}
	dt  
	+ 
	Q^{\hat{x}}_{~u}
	du 
	}^2
	+
	\mathcal{G}_{\hat{u} \hat{u}}
	du^2	
\\
&\quad
	+
	\mathcal{G}_{\hat{t} \hat{x}}
	\roundLR{
	dt
	+
	Q^{\hat{t}}_{~u}
	du 
	}
	\roundLR{
	dx 
	+ 
	Q^{\hat{x}}_{~t}
	dt  
	+ 
	Q^{\hat{x}}_{~u}
	du 
	}
\\
&\quad
	+
	\mathcal{G}_{\hat{t} \hat{u}}
	\roundLR{
	dt
	+
	Q^{\hat{t}}_{~u}
	du 
	}
	du
	+
	\mathcal{G}_{\hat{x} \hat{u}}
	\roundLR{
	dx 
	+ 
	Q^{\hat{x}}_{~t}
	dt  
	+ 
	Q^{\hat{x}}_{~u}
	du 
	}
	du
,
\label{eq:diagonalizedminicoordinatecomponents}
} 
where $Q^{a}_{~b}$ are defined by
\eq{
	Q^{\hat{t}}_{~u}
&=
	\frac{
	{G}_{tx} {G}_{xu}-{G}_{xx}
	   {G}_{tu}
	}{
	{G}_{xt}^2-{G}_{xx} {G}_{tt}
	}
,
\\
	Q^{\hat{x}}_{~t}
&=
	\frac{{G}_{xt}}{{G}_{xx}}
,
\qquad
	Q^{\hat{x}}_{~u}
=
	\frac{{G}_{xu}}{{G}_{xx}}
,
\label{eq:defofQs}
}
and $\mathcal{G}_{\hat{a} \hat{b}}(u)$ is defined as $\mathcal{G}_{\hat{a} \hat{b}}=[(Q^{-1})^{T}GQ^{-1}]_{ab}$ as follows:
\eq{
	\mathcal{G}_{\hat{t} \hat{t}}
&=
	\frac{{G}_{tt}{G}_{xx}-{G}_{xt}^2}{{G}_{xx}}
,
\qquad
	\mathcal{G}_{\hat{x}\hat{x}}
=
	{G}_{xx}
,
\\
	\mathcal{G}_{\hat{u} \hat{u}}
&=
	\frac{
	-\det {G}
	}{
	{g}_{xx}^2
	\roundLR{
	{G}_{xt}^2
	-
	{G}_{tt} 
	{G}_{xx}
	}
	}
.
\label{eq:diagmetric}
}
$\det G$ is the determinant of $G_{ab}$.
${G}_{tt}{G}_{xx}-{G}_{xt}^2$ is of the order of $u-u_{*}$, and vanishes at $u=u_{*}$. 
The other components of $\mathcal{G}_{\hat a \hat b}$ are 
$
	\mathcal{G}_{\hat t \hat x}
=
	O(u-u_{*})
$, $
	\mathcal{G}_{\hat t \hat u}
=
	O(u-u_*)
$, and $
	\mathcal{G}_{\hat x \hat u}
=
	O(1)
$ at $u=u_*$.

The Hawking temperature is read from $\mathcal{G}_{\hat{t} \hat{t}}$ and $\mathcal{G}_{\hat{u} \hat{u}}$ in parallel with (\ref{eq:T}):
\eq{
	T_{*}
=
	\frac{1}{4\pi}
	\sqrt{\frac{a_*}{b_*}}
,
}
where $a_*$ and $b_*$ are defined by
$
	\mathcal{G}_{\hat{t} \hat{t}}
=
	-
	a_*
	(u-u_*)
	+
	O
	((u-u_*)^2)
$ and $
	\mathcal{G}_{\hat{u} \hat{u}}
=
	b_*/(u-u_*)
	+
	O
	(1)
$.
This is the effective temperature $T_{*}$ in the frame of the observer $O_0$.


\begin{thebibliography}{99}

\bibitem{Cugliandolo:2011} 
	L.~F.~Cugliandolo,
	{J. Phys. A: Math. Theor. \bf 44}, 483001 (2011)
	[arXiv:1104.4901 [cond-mat.stat-mech]].

\bibitem{Berthier:2002} 
	L.~Berthier and J.~L.~Barrat,
	Phys. Rev. Lett. {\bf 89}, 095702 (2002)
	[arXiv:cond-mat/0110257 [cond-mat.stat-mech]].

\bibitem{Hayashi:2004} 
	K.~Hayashi and S.~I.~Sasa,
	Phys. Rev. E {\bf 69}, 066119 (2004)
	[arXiv:cond-mat/0309618 [cond-mat.stat-mech]].

\bibitem{Gubser:2006nz} 
	S.~S.~Gubser,
	Nucl.\ Phys.\ B {\bf 790}, 175 (2008)
	[arXiv:hep-th/0612143].  

\bibitem{Nakamura:2013yqa} 
  S.~Nakamura and H.~Ooguri,
  {Phys.\ Rev.\ D \bf 88}, 126003 (2013)
  [arXiv:1309.4089 [hep-th]].

\bibitem{Hoshino:2014nfa} 
  H.~Hoshino and S.~Nakamura,
  {Phys.\ Rev.\ D \bf 91}, 026009 (2015)
  [arXiv:1412.1319 [hep-th]].

\bibitem{Balescu:1968} 
	R.~Balescu, 
	{Physica \bf 40}, 309-338 (1968).

\bibitem{Dunkel:2009tla} 
	J.~Dunkel and P.~H$\ddot{\text{a}}$nggi,
	{Phys. Rept. \bf 471}, 1-73 (2009)
	[arXiv:0812.1996 [cond-mat.stat-mech]].

\bibitem{Farias:2017} 
	C.~Far$\acute{\text{\i}}$as, V.~A.~Pinto, and P.~S.~Moya,
	Scientific reports {\bf 7}, 17657 (2017).

\bibitem{Einstein:1907} 
	A.~Einstein, 
	``$\ddot{\text{U}}$ber das Relativit$\ddot{\text{a}}$tsprinzip und die aus demselben gezogenen Folgerungen," 
	{Jahrbuch Radioaktivit$\ddot{\text{a}}$t Elektron.} {\bf 4}, 411-462 (1907).
	
\bibitem{Planck:1908} 
	M.~Planck,
	{Sitzungsber. Preuss. Akad. Wiss., Berlin}, 542-570 (1907).

\bibitem{Landsberg:1966} 
	P.~T.~Landsberg,
	{P. Phys. Soc. \bf 89}, 1007-1016 (1966).

\bibitem{Kampen:1968} 
	N.~G.~van Kampen,
	{Phys. Rev. \bf 173}, 295-301 (1968).

\bibitem{Ott:1963} 
	H.~Ott,
	{Z. Phys. \bf 175}, 70-104 (1963).

\bibitem{Landsberg:1980} 
	P.~T.~Landsberg,
	Phys.\ Rev.\ Lett.{ \bf 45}, 149 (1980).

\bibitem{Kaniadakis:2002zz} 
  G.~Kaniadakis,
  Phys.\ Rev.\ E {\bf 66}, 056125 (2002)
  [arXiv:cond-mat/0210467 [cond-mat.stat-mech]].
  
\bibitem{Kaniadakis:2005zk} 
  G.~Kaniadakis,
  Phys.\ Rev.\ E {\bf 72}, 036108 (2005)
  [arXiv:cond-mat/0507311 [cond-mat.stat-mech]].
  

\bibitem{Israel:1976tn} 
  W.~Israel,
  Annals Phys.\  {\bf 100}, 310 (1976).


\bibitem{Maldacena:1997re} 
  J.~M.~Maldacena,
  {Int.\ J.\ Theor.\ Phys.\  \bf 38}, 1113-1133 (1999).
  [arXiv:hep-th/9711200].

\bibitem{Gubser:1998bc} 
  S.~S.~Gubser, I.~R.~Klebanov, and A.~M.~Polyakov,
  {Phys.\ Lett.\ B \bf 428}, 105-114 (1998)
  [arXiv:hep-th/9802109].

\bibitem{Witten:1998qj} 
  E.~Witten,
  {Adv.\ Theor.\ Math.\ Phys.\  \bf 2}, 253-291 (1998)
  [arXiv:hep-th/9802150].

\bibitem{Hubeny:2010ry} 
  V.~E.~Hubeny and M.~Rangamani,
  {Adv.\ High Energy Phys.\  \bf 2010}, 297916 (2010)
  [arXiv:1006.3675 [hep-th]].

\bibitem{Gursoy:2010aa} 
  U.~Gursoy, E.~Kiritsis, L.~Mazzanti, and F.~Nitti,
  {JHEP \bf 1012}, 088 (2010)
  [arXiv:1006.3261 [hep-th]].
    
\bibitem{Sonner:2012if} 
  J.~Sonner and A.~G.~Green,
  {Phys.\ Rev.\ Lett.\  \bf 109}, 091601 (2012)
  [arXiv:1203.4908 [cond-mat.str-el]].

\bibitem{Hoshino:2017air} 
  H.~Hoshino and S.~Nakamura,
  {Phys.\ Rev.\ D \bf 96}, 066006 (2017)
  [arXiv:1708.01993 [hep-th]].

    
\bibitem{Juttner:1911} 
	F.~J$\ddot{\text{u}}$ttner,
	{Ann. Phys. \bf 339}, 856-882 (1911).

\bibitem{Cubero:2007} 
	D.~Cubero, J.~Casado-Pascual, J.~Dunkel, P.~Talkner, and P.~H$\ddot{\text{a}}$nggi, 
	{Phys. Rev. Lett. \bf 99}, 170601 (2007)
	[arXiv:0705.3328 [cond-mat.stat-mech]].


\bibitem{Karch:2007pd} 
  A.~Karch and A.~O'Bannon,
  {JHEP \bf 0709}, 024 (2007)
  [arXiv:0705.3870 [hep-th]].


\bibitem{Kim:2011qh} 
  K.~Y.~Kim, J.~P.~Shock, and J.~Tarrio,
  {JHEP \bf 1106}, 017 (2011)
  [arXiv:1103.4581 [hep-th]].

\bibitem{deBoer:2008gu}
J.~de Boer, V.~E.~Hubeny, M.~Rangamani and M.~Shigemori,
JHEP \textbf{07}, 094 (2009)
[arXiv:0812.5112 [hep-th]].

\bibitem{Itzhaki:1998dd} 
  N.~Itzhaki, J.~M.~Maldacena, J.~Sonnenschein, and S.~Yankielowicz,
  {Phys.\ Rev.\ D \bf 58}, 046004 (1998)
  [hep-th/9802042].

\bibitem{Srinivasan:1998ty} 
  K.~Srinivasan and T.~Padmanabhan,
  {Phys.\ Rev.\ D \bf 60}, 024007 (1999)
  [arXiv:gr-qc/9812028].
  
\bibitem{Ammon:2009jt} 
  M.~Ammon, T.~H.~Ngo, and A.~O'Bannon,
  {JHEP \bf 0910}, 027 (2009)
  [arXiv:0908.2625 [hep-th]].
    
\end{thebibliography}
\end{document}